\documentclass[a4paper,12pt]{article}
\usepackage[utf8]{inputenc}
\usepackage{setspace}
\usepackage{graphicx}
\usepackage{color}
\usepackage{subfig}
\usepackage{textcomp}
\usepackage{gensymb}

\def \rarr {\rightarrow}
\def \dstl {\displaystyle}
\def \grinp {\includegraphics}
\def \tw {\textwidth}

\title{Timelike Compton Scattering at JLab}
\author{R.G. Paremuzyan}

\begin{document}


%

\begin{center}
\large
Timelike Compton Scattering at JLab \\
\normalsize
R.G. Paremuzyan$^{1, 2}$
\end{center}

{
\noindent
\footnotesize
1. Institut de Physique Nucléaire, 15 rue Georges Clémenceau, 91406 - ORSAY Cedex - FRANCE \\
2. Yerevan Physics Institute, 2 Alikhanyan Brothers St.,Yerevan 0036, Armenia.
}
\vskip 0.5cm
{

It is demonstrated, that with exclusive final state, data from electron scattering experiments
that are recorded with loose trigger requirements can be used to analyze photoproduction reactions.
A preliminary results on Timelike Compton Scattering using the electroproduction
data from the CLAS detector at Jefferson Lab are presented. 
In particular, using final state ($pe^-e^+$) photoproduction of vector mesons and timelike photon is studied. 
Angular asymmetries in Timelike Compton Scattering region is compared with model predictions
in the framework of Generalized Parton Distribution.
}

\section{Introduction}
In the description of the nucleon structure an important roll plays a formalism of Generalized
Parton Distributions (GPD)s \cite{GPDs}. GPDs provide 3 dimensional description of the quark-gluon structure of the nucleon.
QCD factorization theorem for exclusive processes \cite{Factorize} allows to access GPDs through
exclusive processes in a certain kinematic domain ($Q^{2}>>$, $t/Q^{2} << 1$, $s > 4\;GeV^{2}$).
Theoretically and experimentally best studied reaction in GPD framework is Deeply Virtual Compton Scattering
(DVCS) i.e. $\gamma^{*}p\rightarrow \gamma p$. where incoming photon
has large spacelike virtuality, whereas the outgoing photon is on shell.
GPDs enter into Real part of Compton Form Factors (CFFs) as integral over $x$ (quark internal loop momentum),
or into imaginary part at $x = \pm\xi$ point. The variable $\xi$ (skewdness) is defined as $\Delta^{+} = -2\xi \overline{P}^{+}$, 
where $\Delta = p - p^{\prime}$ is the momentum transfer in the process and $\overline{P} = (p + p^{\prime})/2$
is the average nucleon momentum. Until now most of the published DVCS observables are sensitive to $Im$ part of CFF,
whereas real part is accessible through Beam Charge Asymmetries (BCA)\cite{BCA},
cross-section measurements \cite{DVCS_CRS}, or double spin asymmetries \cite{DSA}.
Measurement of the $Re$ part of CFFs is also important for constraining GPDs, and the $Re$ part of CFFs
have strong model sensitivity, whereas this is less the case for the $Im$ part. This is 
\begin{figure}[!htb]
 \centering
\grinp[width=0.65\tw]{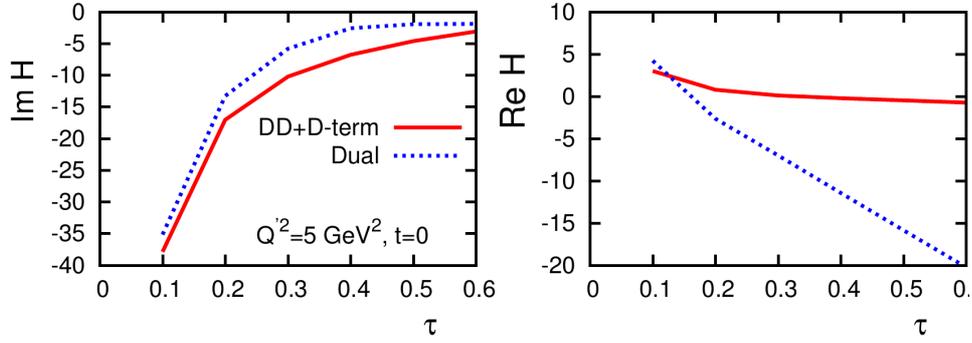}
\caption{$Im$ (left) and $Re$ (right) parts of CFF$\cal{H}$ as a function of $\tau$ 
for DD (solid line), and Dual Parametrization (dotted) models as a function
of $\tau = Q^{\prime 2}/2ME_{\gamma^{*}}$}
\label{fig:CFF_Model_Dep}
\end{figure}
 shown in Fig.\ref{fig:CFF_Model_Dep} where for two models of GPDs, 
double distribution \cite{DD} and dual parametrization \cite{Dual} the $Im$ and $Re$ parts of the CFF $\cal{H}$ are presented.

 Recently a lot of theoretical work has been devoted to the evaluation of the next to leading order (NLO) corrections
 to CFFs, e.g. \cite{NLO1} and \cite{TCS_DVCS_NLO}. As in \cite{TCS} shown, the $Re$ and $Im$ parts of CCFs
can be accessed also through angular asymmetries in the inverse DVCS process, called
Timelike Compton Scattering (TCS), using unpolarized and circularly polarized photon beams, respectively.
 In TCS, $\gamma p \rightarrow \gamma^{*}(\rightarrow l^{-}l^{+})p$ the incoming photon is on shell, 
and the outgoing photon has large timelike virtuality and decays into lepton pairs.
TCS also is an important reaction for testing universalities of GPDs, like in the same way Drell-Yan
was used for checking universalities of PDFs.

\section{Current status of experimental measurements}
\begin{figure}[!htb]
 \centering
 \subfloat[]{\label{fig:quasireal_diagram}\grinp[width=0.3\tw]{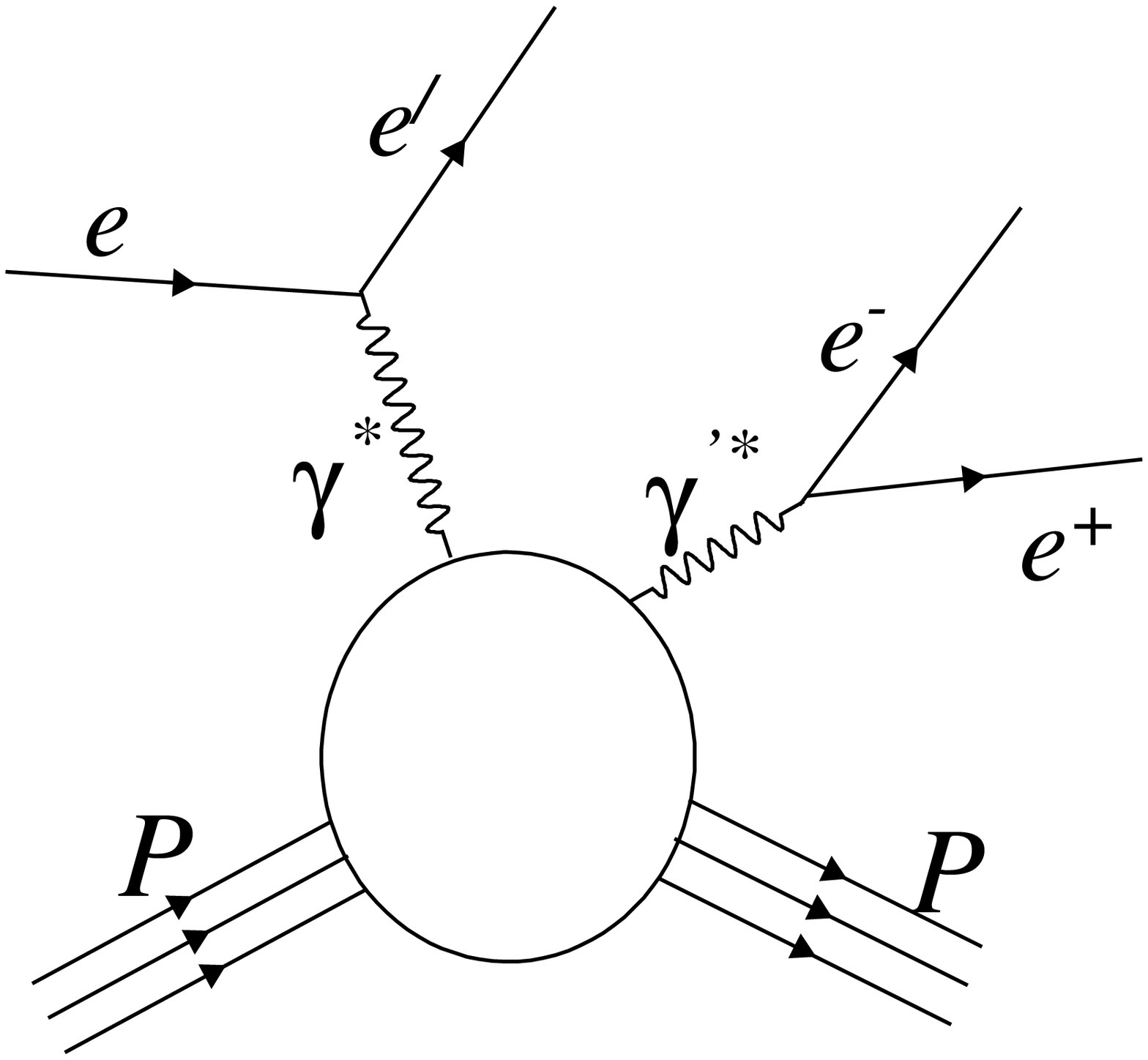}}
 \subfloat[]{\label{fig:Minv}\grinp[width=0.3\tw]{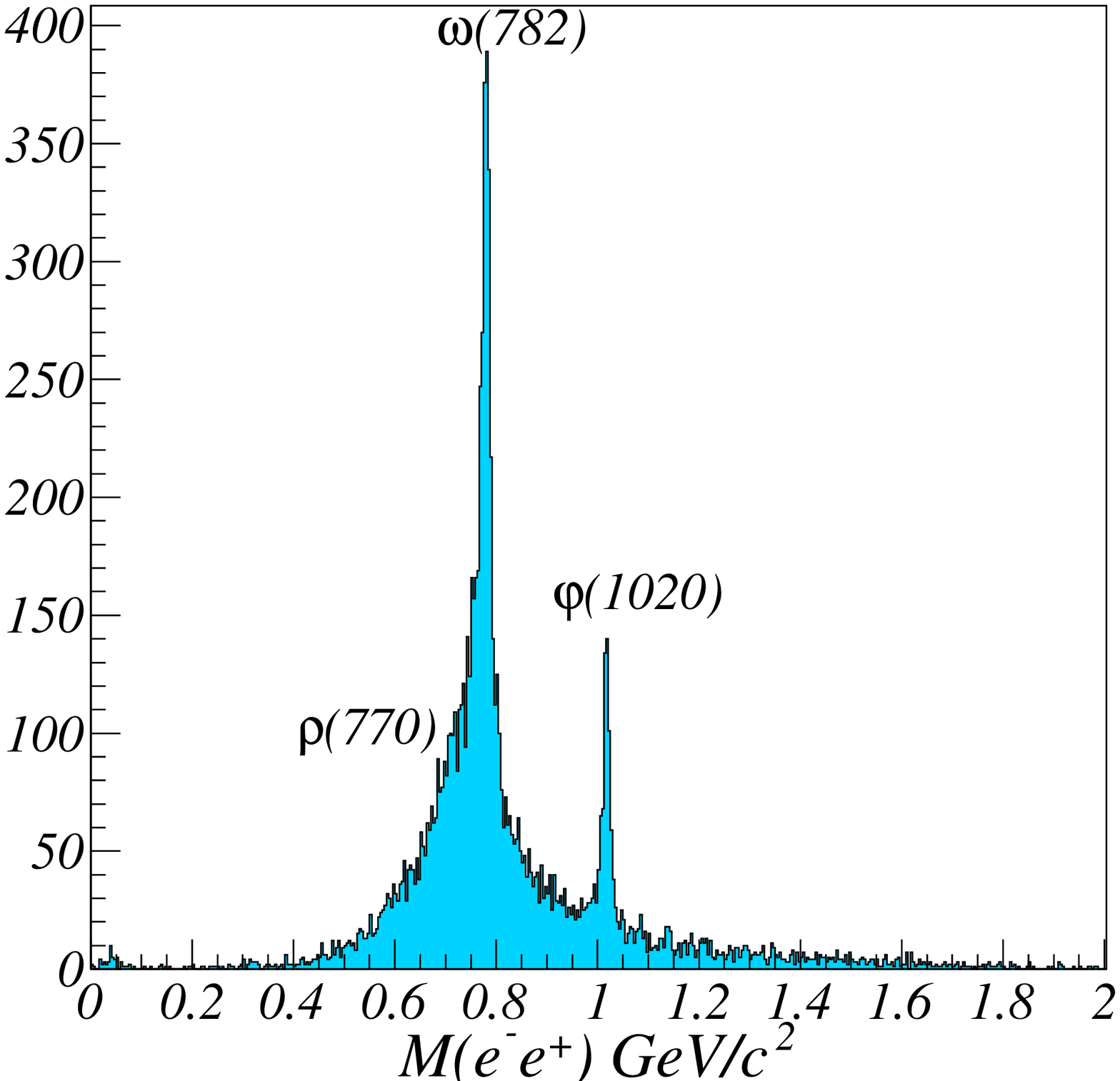}}
 \subfloat[]{\label{fig:R_Data_Theor}\grinp[width = 0.3\tw]{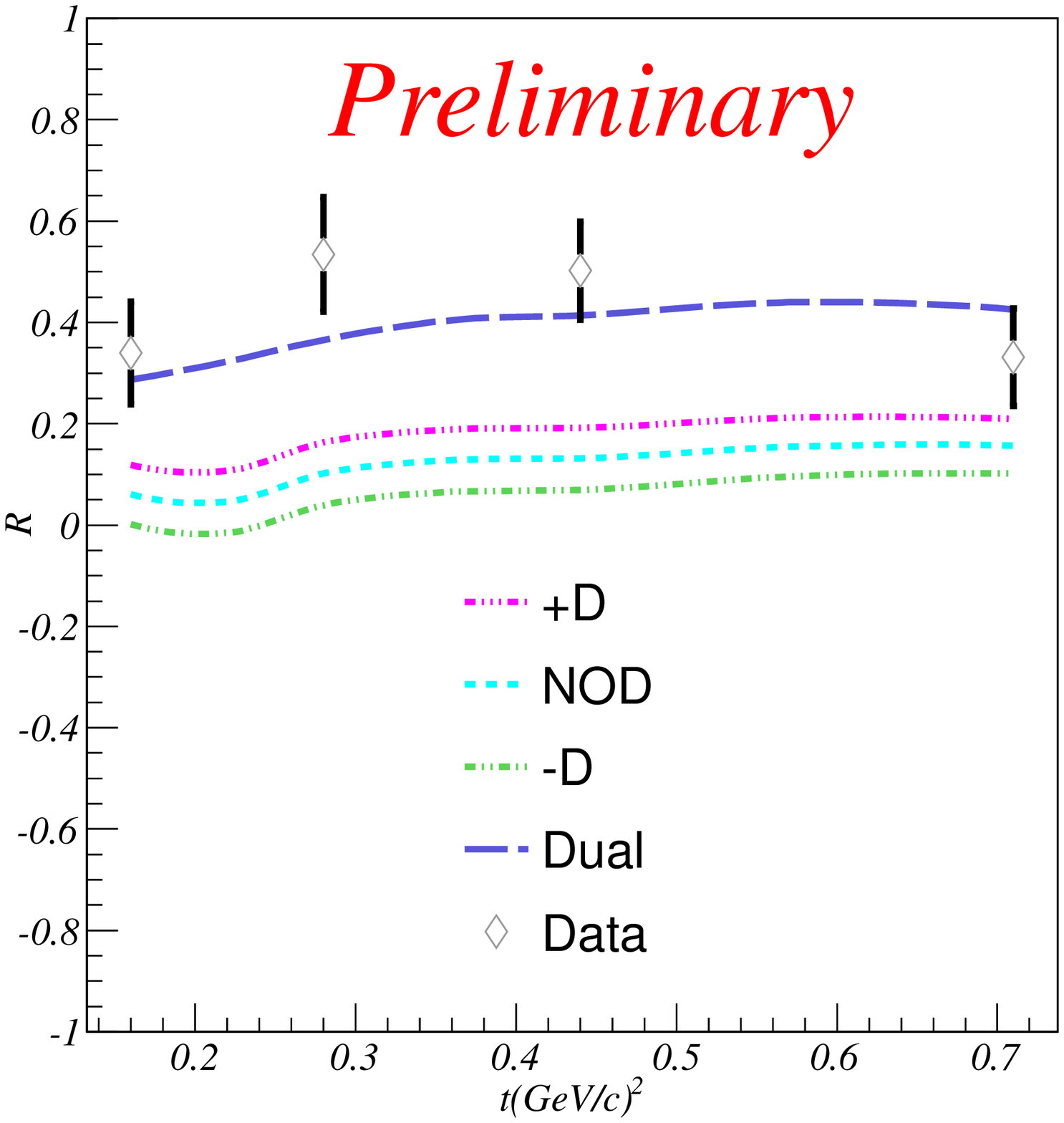}}
 \caption{a) Diagramm for the quasireal photoproduction of lepton pairs., b) Invariant mass of $e^{-}e^{+}$ pairs from data.,
 c) Comparison of the ratio R to theoretical models: The curves $Dual$ represents Dual parametrisation
x model, $+D\;, NOD\;, -D$ represent DD model respectively with Dterm value $+1,\; 0,\; -1$} 
\end{figure}

Experimental data used in this analysis are from two high energy electroproduction experiments 
with the CLAS detector at Jefferson Lab \cite{CLAS} at beam energies $5.76\;GeV$ and $5.479\;GeV$.
In electroproduction experiments, when beam electron scatters at small $\sim 0 \degree$ angle,
the interacting intermediate photon will have $Q^{2}\sim 0$, and 
the reaction $ep\rarr e^{-}e^{+}p(e^{\prime})$ (Fig.\ref{fig:quasireal_diagram}),
where $e^{\prime}$ is the scattered electron, can be interpreted
as quasireal photoproduction of lepton pairs.
The exclusivity of the event is ensured through the 
$Q^{2} < 0.01\;GeV^{2}$ and $|M_{x}|^{2} < 0.1\;GeV^{2}$ cuts, where $|M_{x}|^{2}$ is the missing mass squared of
the $e^{-}e^{+}p$ system and the $Q^{2}$ is calculated from missing momentum. The invariant mass distribution of
$e^{-}e^{+}$ pairs is shown in Fig.\ref{fig:Minv}. One can clearly see peaks for $\omega(782)$
and $\varphi(1020)$ mesons. The sholder on the left side of the $\omega(782)$ corresponds $\rho(770)$. For the TCS analysis
 we have used data above $\varphi(1020)$,
$M(e^{-}e^{+}) > 1.01\;GeV$.
The proposed observable in \cite{TCS} is the normalized ratio R, which is directly related to the
scattering amplitude, and is defined as:
\begin{equation}
 R = \frac{\dstl 2 \int_{0}^{2\pi} d\phi cos(\phi) \frac{dS}{dQ^{2}dtd\phi}}{\dstl \int_{0}^{2\pi} d\phi \frac{dS}{dQ^{2}dtd\phi}}
 \label{eq:R_Def}
\end{equation}
where 
\begin{equation}
 \frac{\dstl dS}{\dstl dQ^{2}dtd\phi} = \int_{\pi/2 - \Delta}^{\pi/2 + \Delta}\frac{\dstl L(\theta, \phi)}{\dstl L_{0}\theta}
  \frac{\dstl d\sigma}{\dstl dQ^{2}dtd\phi}
  \label{eq:dS_Def}
\end{equation}
CLAS acceptance is not symmetric w.r.t. $\theta = \pi/2$.
Therefore in order to compare experimental data with theoretical predictions the integration limits
over $\theta$ in eq.(\ref{eq:dS_Def}) was chosen to fit CLAS acceptance.


In Fig.\ref{fig:R_Data_Theor} preliminary result for the extracted ratio $R$ along
with theoretical predictions are shown.
While $\phi$-dependent $\theta$ integration creates artificial asymmetry, and the ratio $R$ doesn't
reflect the scattering amplitude, it is still usefull information for model comparison and as shown
data favor to Dual parametrization model.

Currently work is in progress to to extract the ratio $R$ in the same way as proposed in \cite{TCS}, through the
extrapolation of experimental data to outside of the CLAS acceptance region.

\section{Future plans}
The 12 GeV upgrade of Jefferson lab will provide better conditions for TCS studies.
The expected beam energy will be $11\;GeV$, which will allow to reach $M(e^{-}e^{+}) < 3.7\;GeV$ region.
The expected luminosity for CLAS12 detector in Hall-B is $\approx 10^{35}cm^{-2}s^{-1}$, which is one order of
magnitude higher than the CLAS maximum luminosity at  $6\;GeV$. Proposal for studying 
TCS and $J/\Psi$ photoproduction near threshold on the proton target with $11\;GeV$ electron beam is already approved
 by JLAb Program Advisory Committee (PAC). It is proposed to study TCS in the mass range
 $2\; GeV < M(e^{-}e^{+})<3\;GeV$ where there
is no contribution from meson resonances and pQCD descibes the ratio 
$R(s) = \frac{\dstl \sigma(e^{-}+e^{+}\rarr hadrons, s)}{\dstl \sigma(e^{-}+e^{+}\rarr \mu^{+}\mu^{-}, s)}$
\cite{PDG}. With $11\;GeV$ high luminosity beams  $J/\Psi$ production near threshold
can be thoroughly studied.
Currently there are no published data for the $J/\Psi$ photoproduction near threshold, and
the $J/\Psi$ production mechanism is not well understood. Measurement
of $J/\Psi$ through it's $J/\Psi\rarr e^{-}e^{+}$ decay channel can be studied with the same technique as TCS, 
and will be an important input for understanding the gluonic form factors of the nucleon.

\section{Summary}
TCS offers a complementary way of studying GPDs, and is an important reaction for testing universalities of GPDs.
Analysis of $6\;GeV$ CLAS data showed feasibility of studying TCS using data from electroproduction experiments.
The same analysis technique will be employed for the TCS analysis with $11\;GeV$ electron beam with CLAS12 detector.
\vskip 0.5cm
This work was supported by the French P2IO laboratory of excellence, and by the Department of Education and
Science of Republic of Armenia, Grant-11-1C015

\end{document}